\documentclass[a4paper]{article}
\usepackage{graphicx}

\begin{document}

\begin{flushright} {\large SAGA-HE-165} \end{flushright}

\vspace{10mm}

\begin{center}
{\LARGE The effective $\omega$-meson mass with a vector field self-interaction}
\end{center}

\vspace{10mm}

\centerline{Katsuaki Sakamoto, Hiroaki Kouno and Akira Hasegawa}
\vspace*{0.015truein}
\centerline{\it Department of Physics, Saga University, Saga 840, Japan}
\vspace*{10pt}
\centerline{and}
\vspace*{10pt}
\centerline{Masahiro Nakano}
\vspace*{0.015truein}
\centerline{\it University of Occupational and Environmental Health,
Kitakyushu 807, Japan}
\baselineskip=10pt

\baselineskip=5mm

\vspace{10mm}
\centerline{\large \bf Abstract}
We calculate the effective mass $m_\omega^*$ of the $\omega$-meson
in nuclear medium 
by the $\sigma$-$\omega$ model with a vector field self-interaction (VFSI). 
We found that the direct and the indirect effects of the VFSI almost cancel
each other and the result is changed only slightly at the normal density. 
The VFSI makes $m^*_\omega$ smaller at lower density and makes $m_\omega^*$
larger at higher density.  
However, both effects are not large. 

\vspace{10mm}

The effective meson mass in nuclear medium is one of the most interesting
issues in intermediate energy physics. 
Much attention has been paid to predicting the meson masses in nuclear medium
from various approaches such as quantum hadrodynamics (QHD) 
\cite{rf:Jean}-\cite{rf:Sakamoto},
the QCD sum rules \cite{rf:Hatsuda,rf:Asakawa} and the quark meson coupling
model \cite{rf:Saito}. 
It is widely accepted that the effective mass of the vector meson
($\omega$ and $\rho$ ) decreases as density increases. 
However, quantitatively, the result is considerably variable in each model.  

In QHD, there are some versions of the Lagrangian to reproduce the properties
of the nuclear matter and finite nuclei. 
One interesting version of them is the $\sigma$-$\omega$ model with the
vector field self-interaction (VFSI). \cite{rf:Bodmer} 
From phenomenological analyses, the VFSI is needed
for a better fit of equation of state at high density. 
(See, e.g., reference \cite{rf:Sugahara}.) 

As usual, this VFSI is regarded as an effective interaction caused
by the quantum corrections. 
This means that the effective potential of the VFSI is added to
the effective potential of the total system. 
On the other hand, the effective meson mass can be defined
as a second derivative of the effective potential with respect to
the meson-field expectation value.  
Therefore, the VFSI just shifts the effective $\omega$-meson mass. 
Naively the VFSI is expected to make the effective $\omega$-meson mass larger,
since the potential of the VFSI has a plus sign. 
Is it true or not? 
In this brief report, 
we study the effective $\omega$-meson mass in nuclear medium by
the $\sigma$-$\omega$ model with the VFSI to answer this question. 

We start from the Lagrangian of the $\sigma$-$\omega$ model
with the meson self-interaction: 
\begin{eqnarray}
L &=& \bar{\psi}(i \gamma_{\mu} \partial^{\mu}-M+g_\sigma
\phi-g_\omega \gamma_{\mu}V^{\mu}) \psi
+\frac{1}{2}\partial_{\mu} \phi \partial^{\mu} \phi
-\frac{1}{2}m_\sigma^2 \phi^2-\frac{1}{4}F_{\mu \nu}F^{\mu \nu}
+U_\omega+\delta L,
\nonumber \\
U_\omega &=&
\frac{1}{2}m_\omega^2V_{\mu}V^{\mu}
\left( 1+\frac{g_\omega^2}{2}Y^2V_{\mu}V^{\mu} \right),
\label{eq:l}
\end{eqnarray}
where $\psi$, $\phi$, $V_{\mu}$, $M$, $m_\sigma$, $m_\omega$,
$g_\sigma$ and $g_\omega$ are the nucleon field, the $\sigma$-meson field,
the $\omega$-meson field, the nucleon mass, the $\sigma$-meson mass, the $\omega$-meson mass, 
the $\sigma$-nucleon coupling and the $\omega$-nucleon coupling, respectively. 
The vector field strength is given by
$F_{\mu \nu}=\partial_{\mu}V_{\nu}-\partial_{\nu}V_{\mu}$, 
and the constant parameter $Y$ represents the strength of
the $\omega$-meson self-interaction.
The $\delta L$ is a counter term adjusted to
satisfy renormalization conditions as in the ordinary relativistic
Hartree approximation (RHA) \cite{rf:Chin}. 
We regard the VFSI as an effective interaction which is originated
in the higher-order quantum corrections beyond the 1-loop approximation,
namely RHA. 

In the RHA, the meson fields are replaced 
by their ground-state expectation values, 
$\phi \rightarrow \phi_0$,
$V_\mu \rightarrow \delta_{\mu 0}V_0$.
The expectation value of the $\omega$ field is determined by
the baryon density $\rho$, \cite{rf:Bodmer} 
\begin{eqnarray}
W=
\frac{g_\omega^2}{m_\omega^2}
\frac{\rho}{1+Y^2W^2}
=
\frac{g_\omega^2}{m_\omega^2}
\frac{\rho}{1+{y^2W^2m_{\omega}^4\over{g_{\omega}^4\rho_0^2}}}, 
\quad W=g_\omega V_0,  
\label{eq:V}
\end{eqnarray}
where $y=g_\omega^2 \rho_0Y/m_\omega^2$ is a dimensionless parameter and
$\rho_0$ is a saturation baryon density. 
The expectation value $\phi_0$ is determined by equation of motion of
the $\sigma$-meson. 
Namely, 
\begin{eqnarray}
g_{\sigma}\rho_s=m_{\sigma}\phi_0, 
\label{eq:ad1}
\end{eqnarray}
where $\rho_s$ is the scalar density of nucleons. 
We remark that the vacuum fluctuation effects is already included
in $\rho_s$ in Eq. (\ref{eq:ad1}). 
The $\phi_0$ shifts the nucleon mass from $M$ to
the effective mass $M^*$ in nuclear medium as 
\begin{eqnarray}
M^*=M-g_{\sigma} \phi_0.
\label{eq:em}
\end{eqnarray}
After $y$ is given, the parameters $C_\sigma =g_\sigma M /m_\sigma$,
$C_\omega=g_\omega M/m_\omega $ are chosen to reproduce the
saturation conditions at the baryon density 0.15 fm$^{-3}$ and
at the binding energy $-$15.75 MeV. 
These two parameters, the effective nucleon mass $M_0^*$
at saturation density and the incompressibility $K$
are listed in Table \ref{ps}. 

\begin{table}[h]
\begin{center}
  \begin{tabular}{lcccccc} \hline \hline
  Model & y & $M_0^*/M$ & $K$ [MeV] & $C_\omega^2$ & $C_\sigma^2$ \\ \hline
  RHA & 0 & 0.730 & 454.1 & 146.86 & 226.94 \\ 
  VS1 & 0.2 & 0.721 & 422.5 & 159.13 & 239.84 \\
  VS2 & 0.5 & 0.692 & 363.4 & 203.59 & 284.73 \\
  \hline \hline
  \end{tabular}
\end{center}
\caption{The parameters in $\sigma$-$\omega$ model.}
\label{ps}
\end{table}

The second order variation of the Lagrangian (\ref{eq:l})
around the classical value $V_0$ gives the inverse of the $\omega$-meson
propagator, 
\begin{eqnarray}
\left(\partial^\mu \partial^\rho
-g^{\mu \rho} \partial_\sigma \partial^\sigma
-\frac{\partial^2 U_\omega}{\partial V_{\mu} \partial V_{\rho}}
\bigg|_{\delta_{\mu 0}V_0} \right)
D(x-y)_{\rho \nu}
= -\delta^\mu_\nu \delta^{(4)}(x-y).
\end{eqnarray}
By taking Fourier transform,
we obtain the inverse of the meson propagator in momentum space, 
\begin{eqnarray}
\left[ D^{-1}(k) \right]_{\mu \nu}
=[k^2-m_\omega^2(1+Y^2W^2)+i\epsilon]g_{\mu \nu}
-k_{\mu}k_{\nu}
-2m_\omega^2Y^2W^2 \delta_{\mu 0} \delta_{\nu 0}.
\end{eqnarray}

In order to compute the $\omega$-meson full propagator in the nuclear medium
we solve Dyson's equation in the random phase approximation (RPA).
Dyson's equation becomes a matrix equation, 
\begin{eqnarray}
D'_{\mu \nu}=D_{\mu \nu}+D_{\mu \rho} \Pi^{\rho \sigma}D'_{\sigma \nu}.
\label{eq:dy}
\end{eqnarray}
We have omitted the $\sigma$-$\omega$ mixing components 
in Eq. (\ref{eq:dy}), since we are interested only in the meson self-energy
with vanishing external spatial momentum as is seen below. 
In the one-loop level, the RPA self-energy of the $\omega$-meson 
is given by
\begin{eqnarray}
\Pi_{\mu\nu}(q) &=& -ig^{2}_{\omega}\int\frac{d^{4}k}{(2\pi)^4}
{\rm Tr}[\gamma_{\mu}G(k)\gamma_{\nu}G(k+q)],
\label{eq:pi}
\end{eqnarray}
where the nucleon propagator $G(k)$ in nuclear medium is given by 
\begin{eqnarray}
G(k) &=& (\gamma^{\mu}k^{\ast}_{\mu}+M^{\ast})
\left[\frac{1}{k^{\ast2}-M^{\ast2}+i\epsilon}
+\frac{i\pi}{E^{\ast}(k)} \delta(k_{0}^{\ast}
-E^{\ast}(k))\theta(k_{F}-|\mbox{\boldmath $k$}|) \right] \nonumber \\
&\equiv& G_{F}(k)+G_{D}(k),
\label{eq:g}
\end{eqnarray}
where $k_F$ is the Fermi momentum,
$k^{\ast\mu} =(k^0-g_\omega V^0,\mbox{\boldmath $k$})$ and
$E^{\ast}(k) = \sqrt{\mbox{\boldmath $k$}^2+M^{\ast 2}}$. 
We remark that the VFSI does not appear in RPA,
since we regard it as an effective interaction induced
by the higher-order quantum correction beyond the 1-loop approximation. 

The equation (\ref{eq:pi}) is divided into two parts
\begin{eqnarray}
\Pi_{\mu \nu}
= \Pi^{\rm Den}_{\mu \nu}+\Pi^{\rm Vac}_{\mu \nu}
=\Pi^{\rm Den}_{\mu \nu}+\left( g^{\mu\nu}-{q^\mu q^\nu\over{q^2}}\right) \Pi^{\rm Vac}. 
\label{eq:pi1} 
\end{eqnarray}
The first term describes particle-hole excitation and the Pauli blocking for
$N$ and $\bar{N}$ excitation below the Fermi surface. 
The second term describes $N$ and $\bar{N}$ excitation,
this part generally has divergence and we remove the divergence
by the renormalization procedure as in \cite{rf:Jean}. 
Namely, 
\begin{eqnarray}
\left.\Pi^{\rm Vac}\right|_{q^2=m_{\omega}^2, \rho =0}&=& 0,
\nonumber\\
\left. \frac{\partial}{\partial q^2}\Pi^{\rm Vac}\right|_{q^2=m_{\omega}^2,\rho =0}
&=& 0.
\label{eq:ad2} 
\end{eqnarray}

The effective meson mass in the nuclear medium is defined as the pole of 
the meson propagator Eq. (\ref{eq:dy}),
namely, given by the following equation, 
\begin{eqnarray}
m_\omega^{* 2}
=m_\omega^2(1+Y^2W^2)
+\Pi_{\rm T}(q_0=m_{\omega}^*,{\mbox{\boldmath $q$}}=0), 
\label{eq:propa}
\end{eqnarray}
where $\Pi_{\rm T}$ is the transverse component of $\omega$-meson self-energy. 
Of course, the term including the parameter $Y$ is occurred
from the existence of the VFSI. 

\begin{figure}
\begin{center}
    \includegraphics*[height=7cm]{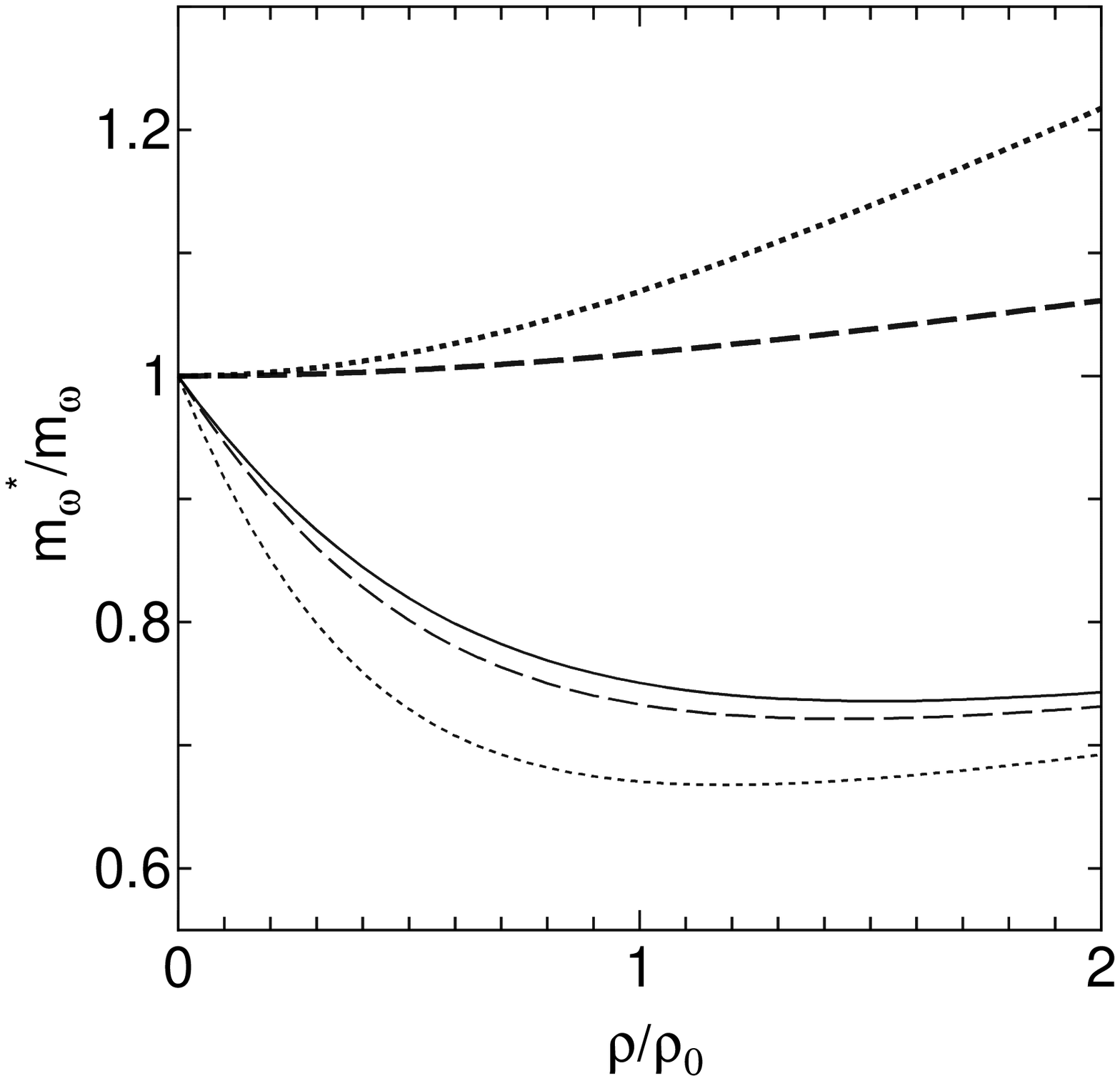}
\caption{
The effective $\omega$-meson mass $m_{\omega}^*$ is shown
as a function of the baryon density $\rho$. 
The bold solid and bold dashed lines correspond to the results for the direct effect with the parameter sets in $y=$ 0.2 and 0.5, respectively. 
The solid, dashed and dotted lines correspond to the results for the indirect effect with the parameter sets in $y=$0, 0.2 and 0.5, respectively.}
\label{mo1}
\end{center}
\end{figure}

First, we examine only the direct effect on the effective $\omega$ meson by adding the VFSI to the effective potential. 
We use the effective $\omega$ meson mass which is defined by 
\begin{eqnarray}
m_\omega^{* 2}
=m_\omega^2(1+Y^2W^2).
\label{eq:mv1}
\end{eqnarray}
In Fig. \ref{mo1}, we show the density dependence of $m_\omega^*$
with Eq. (\ref{eq:mv1}).
The bold solid and bold dashed lines correspond to the results
with the parameter sets in $y=$ 0.2 and 0.5, respectively. 
The $m_\omega^*$ increases as $\rho$ increases. 
The effective potential for the VFSI increases $m_\omega^*$ largely as
is expected. 

Next, we study the VFSI effect on the effective $\omega$-meson mass
via the change of $C_{\omega}$ and $M^*$ in $\Pi_{\rm T}$. 
In this case, we do not contain the direct effect of the VFSI and the effective $\omega$ meson mass is defined by 
\begin{eqnarray}
m_\omega^{* 2}
=m_\omega^2
+\Pi_{\rm T}(q_0=m_{\omega}^*,{\mbox{\boldmath $q$}}=0 \ ; C_\omega,M^*). 
\label{eq:mv2}
\end{eqnarray}
The numerical results of $m_\omega^*$ are also shown in Fig. \ref{mo1}. 
The solid, dashed and dotted lines correspond to the results with the parameter sets for RHA, VS1 and VS2 in table \ref{ps}, respectively.
This "indirect VFSI effect" via the change of $C_{\omega}$ and $M^*$ in $\Pi_{\rm T}$ makes $m_\omega^*$ smaller, since the VFSI makes $C_{\omega}$ larger and makes $M^*_0$ smaller. 

\vspace{0.5cm}
\begin{center}
  \begin{tabular}{c} 
     \hline Fig. \ref{mo1} \\ \hline
  \end{tabular}
\end{center}
\vspace{0.5cm}

\begin{figure}
\begin{center}
    \includegraphics*[height=7cm]{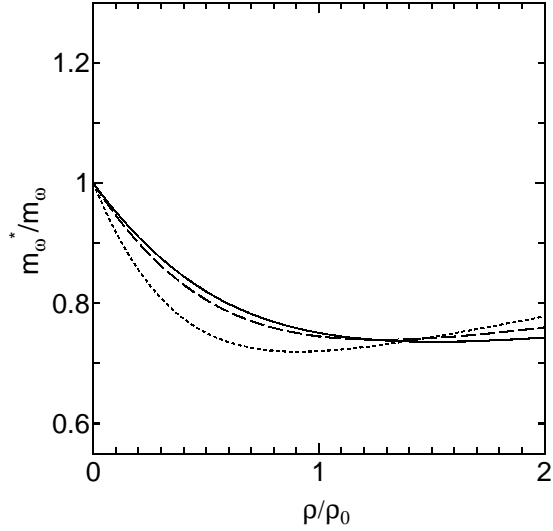}
\caption{
The effective $\omega$-meson mass $m_{\omega}^*$ is shown
as a function of the baryon density $\rho$. 
The solid, dashed and dotted lines correspond to the results with the parameter sets for RHA, VS1 and VS2, respectively. 
}
\label{mo2}
\end{center}
\end{figure}

For a complete calculation in RPA with the VFSI we must use the Eq. (\ref{eq:propa}). 
In Fig. \ref{mo2}, we show the density dependence of $m_\omega^*$ with the Eq. (\ref{eq:propa}). 
We see that the direct and the indirect effects of the VFSI almost cancel each other at the normal density. 
The VFSI makes $m^*_\omega$ smaller at lower density via the large $C_{\omega}$. 
The VFSI makes $m_\omega^*$ larger at higher density due to the direct effect of the VFSI. 
However, both effects are not large. 

\vspace{0.5cm}
\begin{center}
  \begin{tabular}{c} 
     \hline Fig. \ref{mo2} \\ \hline
  \end{tabular}
\end{center}
\vspace{0.5cm}

In summary, we have studied the effect of the VFSI for the effective $\omega$-meson mass in nuclear medium by using the RHA and RPA. 
In the calculation of the RHA, the VFSI increases $C_{\omega}$ to satisfy the saturation conditions. 
It also makes the values of $M^*_0$ and the incompressibility $K$ decrease. 
For the effective $\omega$-meson mass in RPA, 
there is a large direct effect by adding the VFSI to the effective potential. 
However, the direct effect is almost canceled by the indirect effect via the change of $M^*_0$ and $C_{\omega}$ in RPA calculations. 
As a result, the value of the effective $\omega$-meson mass is changed only slightly at the normal density. 
The VFSI makes $m^*_\omega$ smaller at lower density and
makes $m_\omega^*$ larger at higher density. 
However, both effects are not large. 
    
\vspace{10mm}
\begin{center}
{\large \bf Acknowledgements}
\end{center}
$\quad \,$ The authors thank T. Kohmura and T. Maruyama
for useful discussions. 

\vspace{10mm}



\newpage


\end{document}